# Hafnium silicide formation on Si(001)

H. T. Johnson-Steigelman, A.V. Brinck, and P.F. Lyman[a]

*Laboratory for Surface Studies and Department of Physics, University of Wisconsin-Milwaukee,*

*Milwaukee, Wisconsin 53211*

Thick (~50 nm) films of Hf were evaporated onto bare and oxidized Si(001) samples, and thin (~monolayer) films of Hf were evaporated onto clean Si(001)-(2x1) surfaces. Upon annealing to 1000°C, films of $HfSi_2$ were formed after reaction times that depended upon the surface condition of the substrate before deposition. The chemical state of the reacted surfaces were characterized using x-ray photoelectron spectroscopy, and the shifts in binding energy upon silicide formation were recorded. Even for thick films, low-energy electron diffraction (LEED) revealed that the (2x1) pattern of the Si substrate returned, suggesting that 3-dimensional islanding of the $HfSi_2$ film had occurred. The islanding behavior was confirmed for thin films using LEED and x-ray standing waves. Streaking in the LEED patterns for the thick films suggest that the island morphology is influenced by the underlying Si substrate.

---

[a] Corresponding author (plyman@uwm.edu).

**Introduction**

The continued miniaturization of circuit elements in ultra-large scale integrated (ULSI) circuits has required that the $SiO_2$ layer that forms the gate oxide of field-effect transistors be made thinner and thinner. Eventually, a fundamental limit on the thickness of an effective insulating $SiO_2$ barrier will be reached when the layer is so thin that electron tunneling provides a current leakage path that is unacceptably high for low-power devices. This limit is expected to be reached within a few years. As a result, an intensive search has been launched for a material having a higher dielectric constant than $SiO_2$ that could serve as the gate oxide. Such a "high-$\kappa$" insulator would allow the gate oxide to be physically thicker (for a given capacitance), dramatically reducing tunneling. However, to serve as an effective gate insulator, any replacement for $SiO_2$ must possess the many favorable electrical and structural properties of the $SiO_2$/Si interface. In particular, it is thought that oxides and silicates of Hf and Zr may exhibit an interface with Si that is stable against deleterious $SiO_2$ formation, but would possess many of the desirable interface attributes of $SiO_2$ [1, 2].

Hf forms stable silicates $HfSi_xO_y$ with a range of Hf/Si ratios [2], and it has been shown that a relatively small amount of Hf can significantly increase the dielectric constant of $SiO_2$ [3, 4]. Moreover, it has been found that hafnium silicates with a small Hf/Si ratio readily form stable amorphous compounds [5], which are desirable in a gate oxide to reduce electrical leakage and dopant diffusion along grain boundaries. To provide the smallest deviation from the mature techniques established for device construction, it makes sense to produce hafnium silicate films with a small Hf/Si ratio. The present paper reports our investigation of the formation of hafnium silicide films, which could then be oxidized to form a hafnium silicate [6]. To minimize the Hf/Si ratio, we have concentrated on the formation of the most Si-rich stable silicide phase, $HfSi_2$.

The Hf/Si system has been investigated previously in the hope of forming silicide device interconnects having low sheet resistance and contact resistivity [6, 7, 8]. In these works, Hf was deposited on Si substrates, and the kinetics of the reaction to form HfSi and then $HfSi_2$ were monitored. Notable findings were that HfSi and $HfSi_2$ formed readily at temperatures as low as 600°C and 765°C, respectively. Si was found to be the dominant diffusing species. For a film of Hf on crystalline Si, HfSi would form in a layer-wise fashion between the Si substrate and the unreacted Hf, but the formation of $HfSi_2$ would nucleate heterogeneously [7] unless excess Si was supplied in the form of an amorphous Si capping layer [8]. Later work focused on the thin (~12 nm) amorphous hafnium silicide layer that is formed at the Hf/Si interface at annealing temperatures of 400-500°C [9, 10, 11].

In the present work, we report results on the formation of $HfSi_2$ by the solid-state reaction of metallic Hf with Si(001) substrates. We also investigated the effect on this reaction of a layer of $SiO_2$ at the interface between the deposited Hf film and the substrate. The chemical state and compound formation was probed using x-ray photoelectron spectroscopy (XPS) and powder x-ray diffraction (XRD), while important inferences about the morphology of the reacted film were derived from low-energy electron diffraction (LEED) and x-ray standing waves (XSW). Because our primary interest is on the eventual production of hafnium silicate films with a small Hf/Si ratio, this study emphasized the formation of $HfSi_2$ over the more Hf-rich phases formed at lower temperatures.

**Experiment**

Hafnium silicide formation was studied in two regimes: thick films (~50 nm) and thin films (~monolayer, ML). For x-ray photoelectron spectroscopy (XPS) experiments, the substrates consisted of commercial Si(001) wafers (*p*-type, 5 Ω-cm). For the x-ray standing wave (XSW) experiments, the float-zone Si(001) samples were Syton-polished and chemically cleaned *ex situ* using the Shiraki process [12], and then mounted in a strain-free manner.

Samples were analyzed by XPS and low energy electron diffraction (LEED) in a VG Mk. II ESCALab ultrahigh vacuum (UHV) chamber with a base pressure of less than $1\times10^{-10}$ torr. Unmonochromated Al $K_\alpha$ radiation was used to excite the photoelectrons, and spectra were acquired with near-normal electron emission. In addition to a survey spectrum, detailed scans (instrumental resolution ≤ 0.95 eV) of the O 1*s*, C 1*s*, Si 2*p*, and Hf 4*f* regions were acquired for each surface condition. Binding energies were referenced to the Ag $3d_{5/2}$ core level, assumed to lie at 368.27 eV, and were determined with an overall estimated precision of ± 0.05 eV.

Thick Films

For the thick film studies, Si(001) wafers with three different surface oxidation conditions were prepared prior to Hf deposition. One set of samples was placed in a box furnace at 900°C for 6 hours to form a ~50 nm thermal oxide. Another set of samples was used as-received, with a ~1.5 nm native oxide. The third set of samples was stripped of its native oxide using dilute HF acid immediately before insertion into a high vacuum (HV) electron-beam evaporator. Hf was deposited at ~$10^5$ torr on the three types of substrate samples, which were held at RT. Stylus profilometry indicated that the deposited Hf layers were 50 to 80 nm thick. These samples were then transferred through air to the UHV chamber for annealing and analysis. During prolonged annealing of samples to 1000°C, the pressure of the UHV chamber rose from its base pressure to ~$2\times10^{-9}$ torr.

Thin Films

For the thin film studies, samples were inserted into the UHV chamber as received, were degassed, and then resistively heated to 1000°C, rendering a clean (2x1)-reconstructed surface. Submonolayer to ML amounts of 99.9% pure Hf (excluding ~3% Zr) were deposited on the substrates using an electrostatic electron beam evaporator [13]. The samples were held at room temperature (RT), and the pressure typically rose to ~$5\times10^{-9}$ torr during deposition. After deposition and after each subsequent anneal, XPS and LEED measurements were repeated.

The x-ray standing waves (XSW) measurements were conducted at the BESSRC undulator beamline 12ID-D of the Advanced Photon Source (APS) at Argonne National Laboratory. A high-heat-load Si(111) monochromator was used to select a 11.5 to 12.5 keV beam from the undulator output, and then the beam was further monochromated and dispersion-matched to the (004) reflection of the Si(001) substrate using a Si(004) 2-bounce channel-cut crystal. An energy-dispersive Si(Li) detector was used to detect the Hf fluorescent x-ray signal. The XSW technique has been reviewed and the experimental arrangement used has been described by Zegenhagen [14].

**Results**

Thick Films

Before annealing, no Si features appeared in the XPS spectrum. The only Hf feature evident stemmed from $HfO_2$ (from the native oxidation of the deposited Hf films) [15]. Although it had been reported long ago [7] that Hf and Si react at temperatures as low as 600°C to form HfSi and at 765°C to form $HfSi_2$, little change in any of our samples, either visually or by XPS, was observed upon annealing to temperatures less than 950°C. (However, XPS is only sensitive to the outermost several nm.) The only change observed at these temperatures was a reduction of the $HfO_2$ thickness in some samples evidenced by the introduction a metallic feature in the Hf 4$f$ spectra. No spots appeared in the LEED pattern from these samples, indicating that the deposited surface layer lacked long-range order.

Upon annealing to 1000°C, however, the visual appearance of all samples changed dramatically, from dark metallic silver to a matte gray. Importantly, however, the amount of time that must elapse at this annealing temperature before this visual change occurs varied greatly with substrate surface oxidation condition. Specifically, the HF-etched samples reacted in ~10 minutes, the samples having a native oxide took ~1-2 hours, and the thermally oxidized samples took from 3 to 12 hours at 1000°C. Thus, $SiO_2$ appears to form a barrier against the interdiffusion necessary for this reaction to proceed.

For all of the samples, the condition after heating to 1000°C was similar. After annealing (with the concomitant change in visual appearance), all samples gave rise to essentially the same XP spectra; specifically, the Hf 4$f$ region contained only one sharp spin-orbit doublet, and Si was apparent, as shown in Fig. 1. The Hf 4$f$ doublet was extremely narrow, with a linewidth of $\leq 0.95$ eV, and the BE of the $4f_{7/2}$ feature was 14.65 eV. For comparison, we determined the BE of metallic Hf to be 14.35 eV, in good agreement with published reports of 14.31 eV [15] and 14.3 eV [16]. Metallic Hf shows a pronounced asymmetry in the XPS lineshape; upon conversion to $HfSi_2$, the asymmetry of the Hf 4$f$ lineshape was greatly reduced. (The Doniach-Sunjic asymmetry parameter [17] decreased from 0.32 to 0.14 upon silicidation.) The Si $2p_{3/2}$ line exhibited a prominent feature at 99.45 to 99.50 eV. [See Fig. 1(b).] This peak is essentially indistinguishable from that of substrate Si, which we determined to have a Si $2p_{3/2}$ BE of 99.5 eV [18]. Notable by its absence in XPS, even for samples initially containing a 50 nm $SiO_2$ layer, was any feature in the O region. Remarkably, samples annealed at 1000°C for several hours gave rise to a (2x1) LEED pattern reminiscent of the clean surface of Si(001). (See Fig. 2.) As discussed below, these results indicate that 3-dimensional islands of $HfSi_2$ have been formed upon annealing. Subsequent powder x-ray diffraction (XRD) confirmed that annealed samples were completely or dominantly composed of $HfSi_2$. For many of the samples that underwent prolonged annealing and resultant islanding, a Si $2p_{3/2}$ feature at 101.2 eV BE attributable to SiC became evident, as shown in Fig. 3.

Thin Films

Upon RT deposition of ~1 ML Hf onto a clean Si(001)-(2x1) surface, no spots were visible in the LEED pattern, indicating a complete destruction of the surface's long range order by the reactive metal film. The Hf $4f$ spectrum showed a single broad feature centered near 16.0 eV BE (which includes contributions from the $4f_{7/2}$ and $4f_{5/2}$ lines). The feature lacked the asymmetric tail characteristic of metallic Hf. It is not possible to uniquely decompose this peak, but it can be accounted for either as one Hf $4f$ silicide doublet with twice the usual width, or as a combination of several doublets each having the silicide lineshape and slightly displaced BE. The result is consistent with an inhomogeneous layer of Hf atoms bound to Si, but not fully reacted. The Si $2p$ spectra did not reveal any surface components attributable to Hf bonding.

However, as demonstrated with the thick films, little BE shift is expected upon reaction with Hf, and moreover the data were not acquired in a surface-sensitive mode.

Upon annealing to ~750°C for 5 minutes, the Hf 4f spectral features coalesced into a sharp doublet characteristic of the silicide phase, as shown in Fig. 4. The BE of the $4f_{7/2}$ peak was 14.6 eV, which is about 0.6 eV shallower than prior to annealing. Moreover, the LEED pattern changed back to a (2x1) reminiscent of the clean surface (albeit with higher background and broadened features). This process was repeated several times, with additional Hf metal being deposited at RT, which removed the LEED pattern and broadened the XPS features; upon reannealing, the (2x1) pattern reemerged each time and the XPS features sharpened into a silicide-like lineshape with a $4f_{7/2}$ component at 14.6-14.7 eV BE. These observations suggest that the Hf, upon annealing, gathered into islands on the surface, exposing the substrate beneath.

XSW analysis was carried out on thin (~ML) Hf/Si(001)-(2x1) samples. As in the XPS experiments, ML quantities of Hf were deposited on clean Si(001)-(2x1) substrates held at RT using an electrostatic electron beam evaporator. After characterization by LEED and Auger electron spectroscopy (AES), the samples were annealed to 700°C for 2 min. Typical XSW results for such a film are shown in Fig. 5. The results are puzzling at first inspection because the modulation in the normalized fluorescence yield across the substrate rocking curve is smaller than the changes in x-ray reflectivity. As will be more fully explained in the Discussion section, this result can be explained if the HfSi$_2$ forms clusters that are comparable in size with the absorption length in the deposited material of the incident x-rays.

**Discussion**

Evidently, Hf and the Si substrate are able to react at 1000°C despite the presence of a SiO$_2$ layer at the interface with a thickness of up to 50 nm.  The lack of O in the ~1000 °C annealed samples suggests that O that was originally at the interface of the deposited layer and the substrate was transported to the surface and desorbed.  As in the thermal desorption of SiO$_2$, the desorbing species is likely SiO.  This conclusion can explain why the time required to completely react the surface increased with the thickness of the substrate oxide layer.  (Of course, it is also possible that a silicate layer containing O remains at the interface; however, in other studies conducted in our laboratory [1, 19], hafnium silicate layers were observed to break down to HfSi$_2$ at annealing temperatures of 1000 °C.)

After formation of the silicide phase, there are a number of observations that imply that 3-dimensional clusters or islands occur on the surface of both the thin and the thick samples.  For thick films annealed for several hours in UHV, a LEED pattern characteristic of the clean surface appears.  The appearance of this pattern from a ~80 nm film is difficult to explain except if the HfSi$_2$ film forms 3-dimensional structures, thus allowing a fraction of the bare surface to be exposed.  A few samples (i.e., those with no oxide at the Hf/Si interface) reacted and changed their visual appearance in ~10 min.  These samples did not display an ordered LEED pattern after this brief anneal. Moreover, for samples that underwent prolonged annealing, a second feature attributable to SiC appeared at 101.2 eV BE in the Si 2*p* XP spectra. We propose that SiC forms at the bare patches of Si exposed to the residual gases, which is not unexpected for these annealing times, temperatures, and background pressures.

The final piece of evidence for island formation comes from analysis of the XSW results. The results are puzzling at first inspection. For a thin adsorbate layer, the angular dependence of the fluorescence yield $Y(\theta)$ (normalized to its intensity away from the Bragg reflection) can always be expressed as

$$Y(\theta) = 1 + R(\theta) + 2\sqrt{R(\theta)}\, f_\mathbf{H} \cos[\nu(\theta) - 2\pi P_\mathbf{H}] \quad , \tag{1}$$

where $R(\theta)$ is the x-ray reflectivity and $\nu(\theta)$ is the relative phase of the diffracted x-ray plane wave [14]. The coherent fraction $f_\mathbf{H}$ and coherent position $P_\mathbf{H}$ correspond to the amplitude and phase, respectively, of the Hth Fourier component of the time-averaged spatial distribution of the nuclei of the adatoms (projected into a unit cell). **H** is the reciprocal lattice vector for the (hkl) diffraction planes, which was (004) in the present case. For adsorbates or films whose projection into one unit cell bear no correlation with the substrate lattice planes, the coherent fraction $f_\mathbf{H}$ is zero. That value is expected for either a disordered layer, or even for an epitaxial film of more than a few layers thick if the out-of-plane lattice constant is different from that of the reflecting substrate. These facts imply that (for a thin adsorbate layer), even a disordered layer will give rise to a modulation of at least $R$, which is 93% in this case [20]. In Fig. 5, the modulation is only ~50%. However, Eq. 1 only holds for adsorbate layers that are thin compared to the absorption length of the incident x-rays in the adsorbate material. In particular, if the $HfSi_2$ forms large clusters, then the fluorescence yield will not obey Eq. 1. Instead, for a thick film whose atomic positions are not correlated to the substrate, the fluorescence will follow

$$Y(\theta) = 1 + \alpha R(\theta) \quad , \tag{2}$$

where $\alpha$ is related to the fraction of the film that is transparent to the incident x-rays.

For the incident x-ray energies used, the absorption length in $HfSi_2$ is 6.3 to 7.9 µm. In Fig. 5, the best fit to the data was obtained using a value of $\alpha$ of 0.50. We can conclude that the islands that form have a characteristic length scale of µm. (It is not possible to say more than that without knowing the aspect ratio and size distribution of the islands.) Recall that these results were obtained on films of several ML initial Hf coverage. Thus, the tendency to island must be quite pronounced. Many other transition metal silicides form 3-dimensional clusters on Si surfaces [21]. Although the islanding behavior seen here for hafnium silicides is not surprising, it has not been previously reported to our knowledge.

For the thick films, we may infer a little of the morphology of the reacted layer by examining the LEED pattern shown in Fig. 2. Streaking along the (00) to {01} –like and {10} to {11}-like directions indicates that the exposed patches of substrate are long (compared to the electron coherence length) in one dimension, and short in the other. Thus, the exposed substrate patches must be long and narrow, and must be influenced by the substrate crystallographic template.

Unfortunately, the change in surface morphology (i.e., islanding) observed upon high-temperature reaction is unfavorable for the intended eventual formation of a uniform hafnium silicate layer having a small Hf/Si ratio by oxidation of a $HfSi_2$ layer. However, no LEED spots were observed from the samples that completely reacted in 10 min., so it may be possible to limit the kinetics of the island formation using, e.g., rapid thermal processing.

The BE of the Hf and Si core levels change little upon reaction to form $HfSi_2$. It is tempting to try to interpret BE shifts in XPS as arising (primarily) from charge transfer between atoms. In the present case, the Hf $4f_{7/2}$ line shifts 0.35 eV to deeper BE, while the Si $2p_{3/2}$ line barely shifts at all. If one attempts to interpret these shifts as stemming from charge transfer, it is difficult to understand why so little BE shifting occurs. One expects that Hf, which has a Pauling electronegativity of 1.3, would donate charge to Si, which has a Pauling electronegativity of 1.9; this should result in a substantial shift to deeper BE for Hf and a shift to shallower BE for Si. However, this simple model is not reliable. In particular, it has been shown that it is unable to account for BE shifts upon alloy and compound formation [22]. Our results for BE shifts are similar overall to the findings of Zaima *et al*. [10], but differ in detail. They report a Hf $4f_{7/2}$ BE shift from 13.8 eV (for the deposited Hf metal) to 14.1 eV upon formation of $HfSi_2$, which (except for a 0.5 eV absolute discrepancy) is consistent with our reported shift. However, they report a Si $2p$ BE shift from 99.3 eV for bulk Si to 99.1 eV after reaction, presumably to $HfSi_2$. We do not observe this 0.2 eV shift to shallower BE for the Si core level.

**Conclusions**

Hf reacts with Si(001) to form $HfSi_2$ at 1000°C, regardless of any initial $SiO_2$ layer on the Si substrate. An interposing $SiO_2$ layer, however, will considerably slow the kinetics of the reaction. After annealing, O is removed, probably having been desorbed as SiO. The resultant $HfSi_2$ exhibits a sharp, fairly symmetric XPS lineshape with BE of the $4f_{7/2}$ core level of 14.65 eV, while the Si $2p$ levels exhibit negligible BE shift upon reaction. The resultant film exhibits a pronounced tendency to form 3-dimensional islands, leaving patches of the substrate denuded of Hf. The shape of the bare patches is long and narrow, indicating that the underlying crystal structure influences the formation of the islands.

**Acknowledgements**

This material is based on work supported by the National Science Foundation under Award

Number DMR-9984442 and by an REU Site Award. Use of the Advanced Photon Source was supported by the U. S. Department of Energy, Office of Science, Office of Basic Energy Sciences, under Contract No. W-31-109-Eng-38. Use of the facilities of BESSRC is gratefully acknowledged. The authors are grateful to Daniel Shillinglaw for assistance in data collection and to Michael Weinert for useful discussions.

**Figure Captions**

Fig. 1 (a) Hf 4*f* XPS spectrum from thick (~50 nm) $HfSi_2$ sample (bottom) and clean metal Hf foil (top). (b) Si 2*p* spectrum from thick $HfSi_2$ sample (bottom) and clean Si(001)-(2x1) (top). The $HfSi_2$ sample was formed on HF-etched Si(001) and had been annealed to 1000°C for 50 min. A Shirley background was removed prior to fitting. Spectra are scaled to the same height and are offset for clarity.

Fig. 2. LEED pattern acquired at 70 eV from thick (~50 nm) $HfSi_2$ film on HF-etched Si(001) after annealing at 1000°C for 3 hours. For clarity of presentation, the video image has been inverted.

Fig. 3. Si 2*p* XPS spectrum and fit acquired from thick (~50 nm) $HfSi_2$ sample on thermally oxidized Si(001) after annealing at 1000°C for 12 hours. A Shirley background was subtracted before fitting. The long (short) dashed lines correspond to the $HfSi_2$ (SiC) components.

Fig. 4. Hf 4*f* XPS spectra acquired from thin (~ML) Hf deposited at RT on clean Si(001)-(2x1) before and after annealing to 750°C for 5 min. The lines are a guide to the eye.

Fig. 5 XSW data from thin (several ML) Hf deposited at RT on clean Si(001)-(2x1) and then annealed to 700°C for 2 min. The lower data are the x-ray reflectivity from the Si(004)

reflection, and the curve is a fit using dynamical diffraction theory. The upper data are the Hf $L_\beta$ fluorescent yield (normalized to the off-Bragg value), and the curve is a fit to Eq. 2, with $\alpha=0.50$.

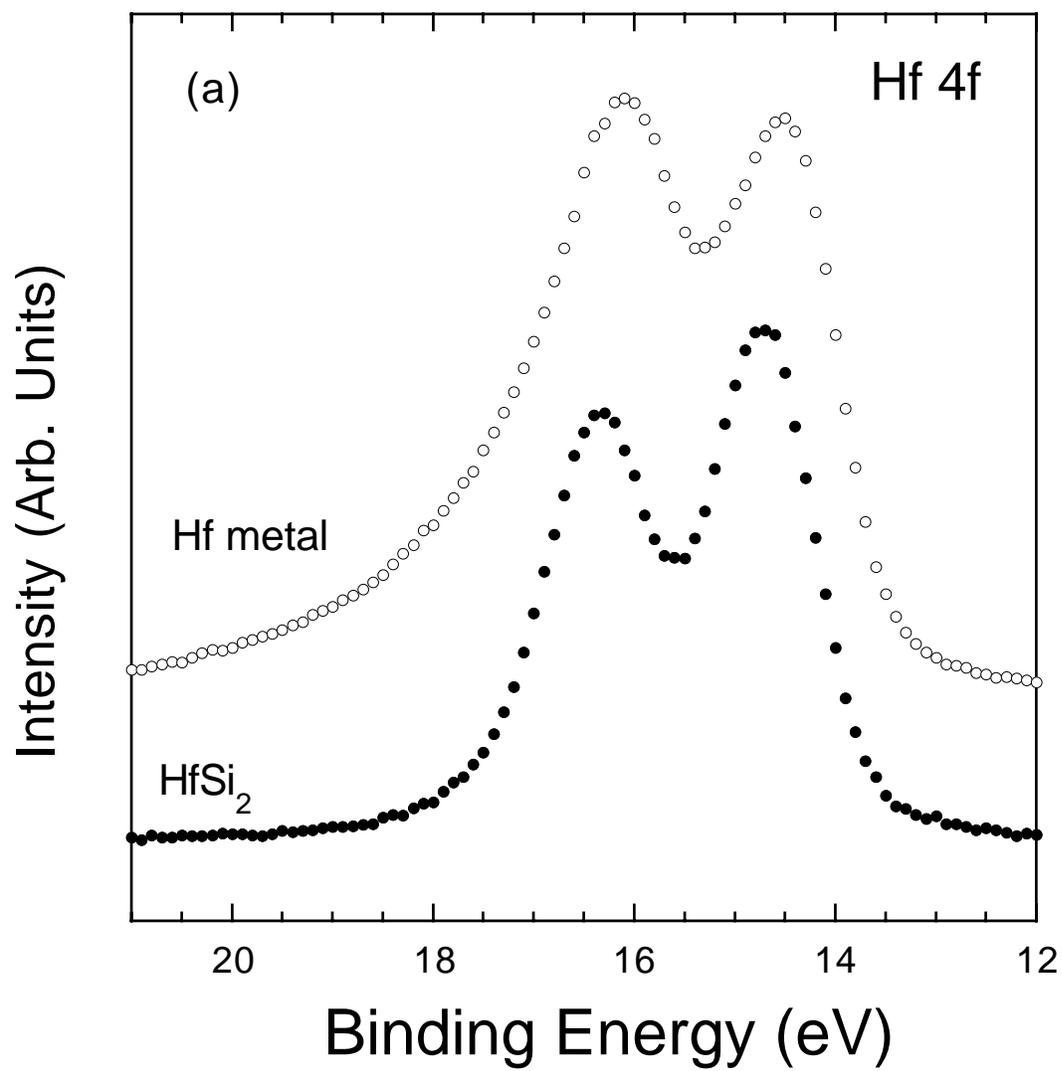

Fig. 1(a) Johnson-Steigelman et al.

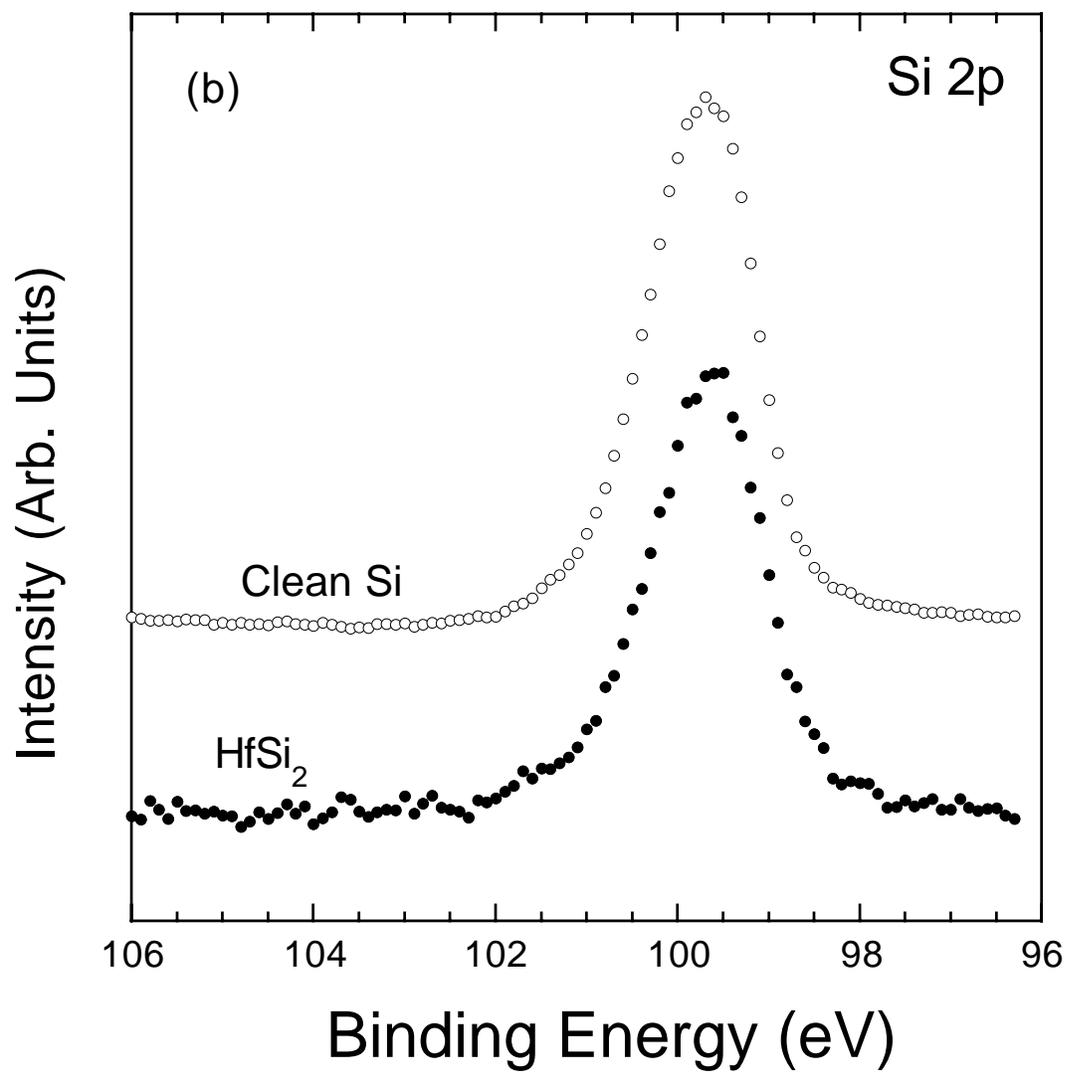

Fig. 1(b)  Johnson-Steigelman et al.

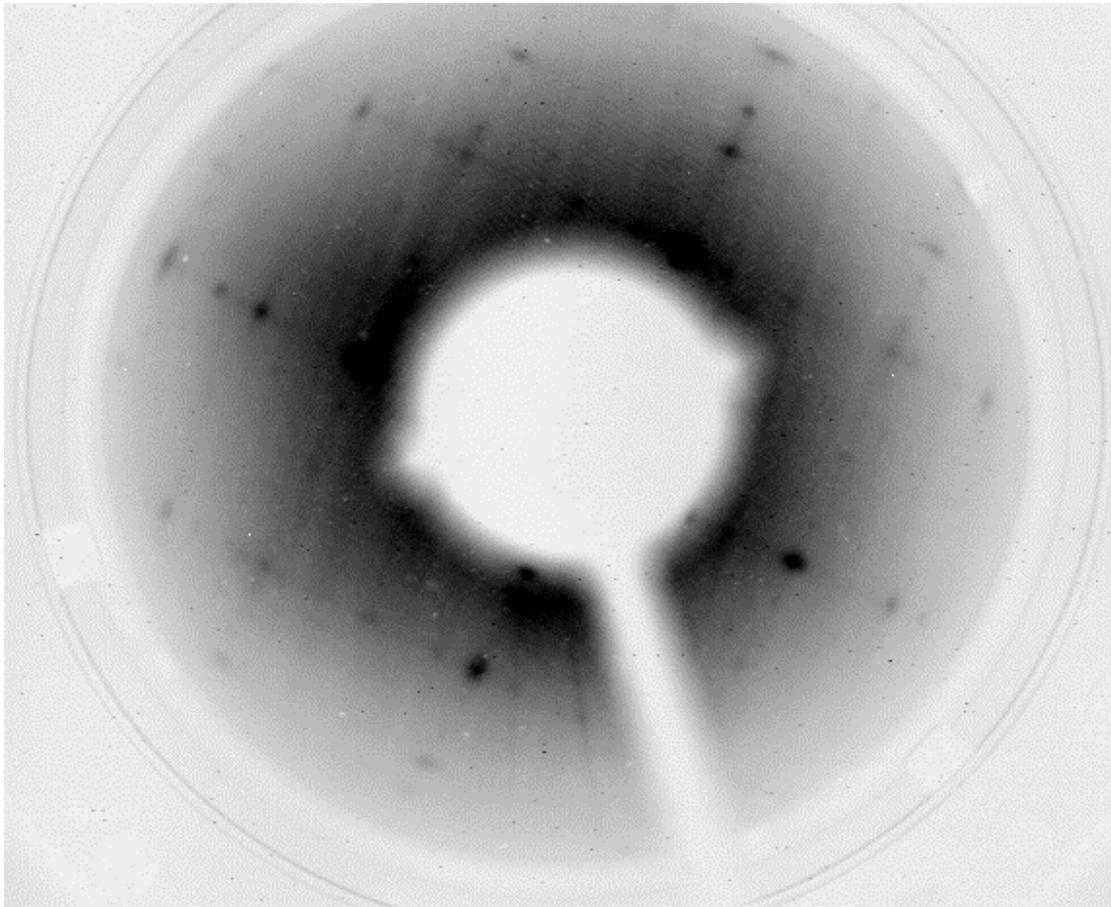

Fig. 2  Johnson-Steigelman et al.

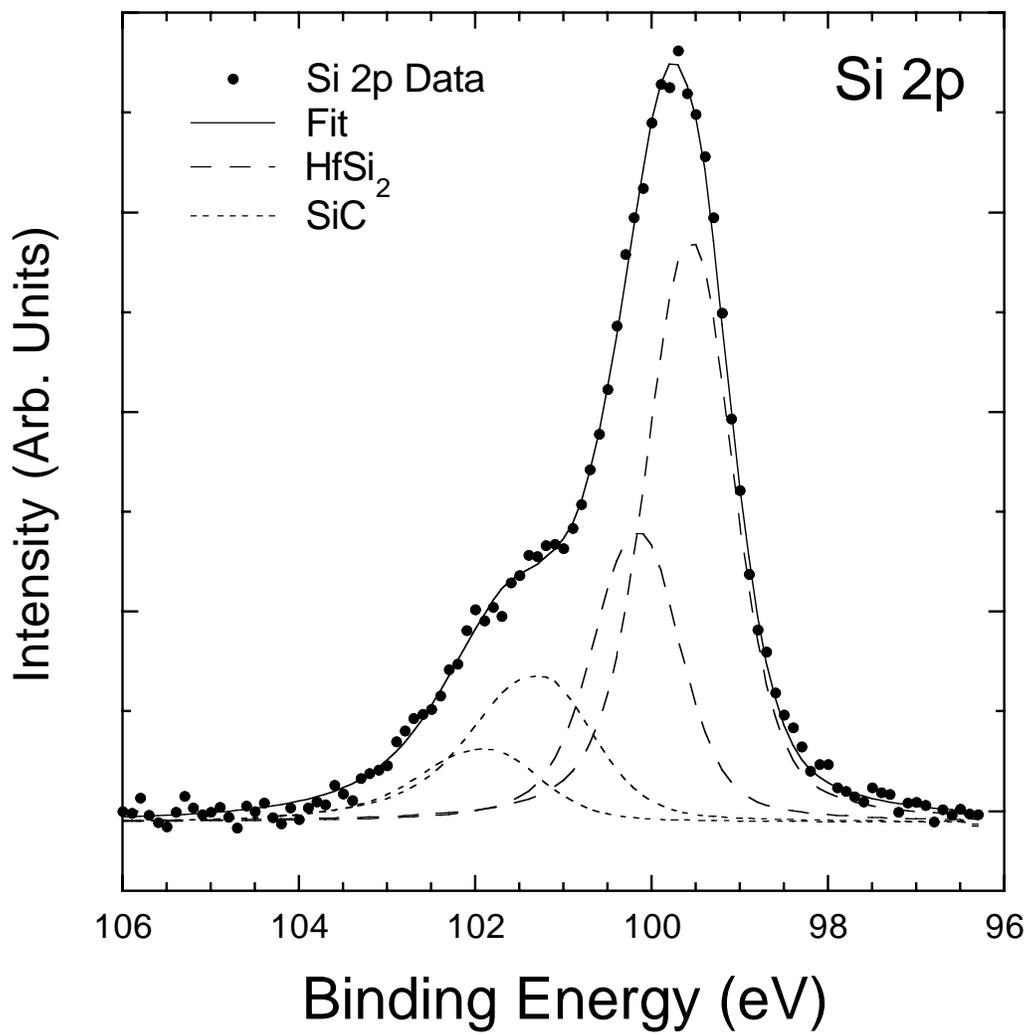

Fig. 3  Johnson-Steigelman et al.

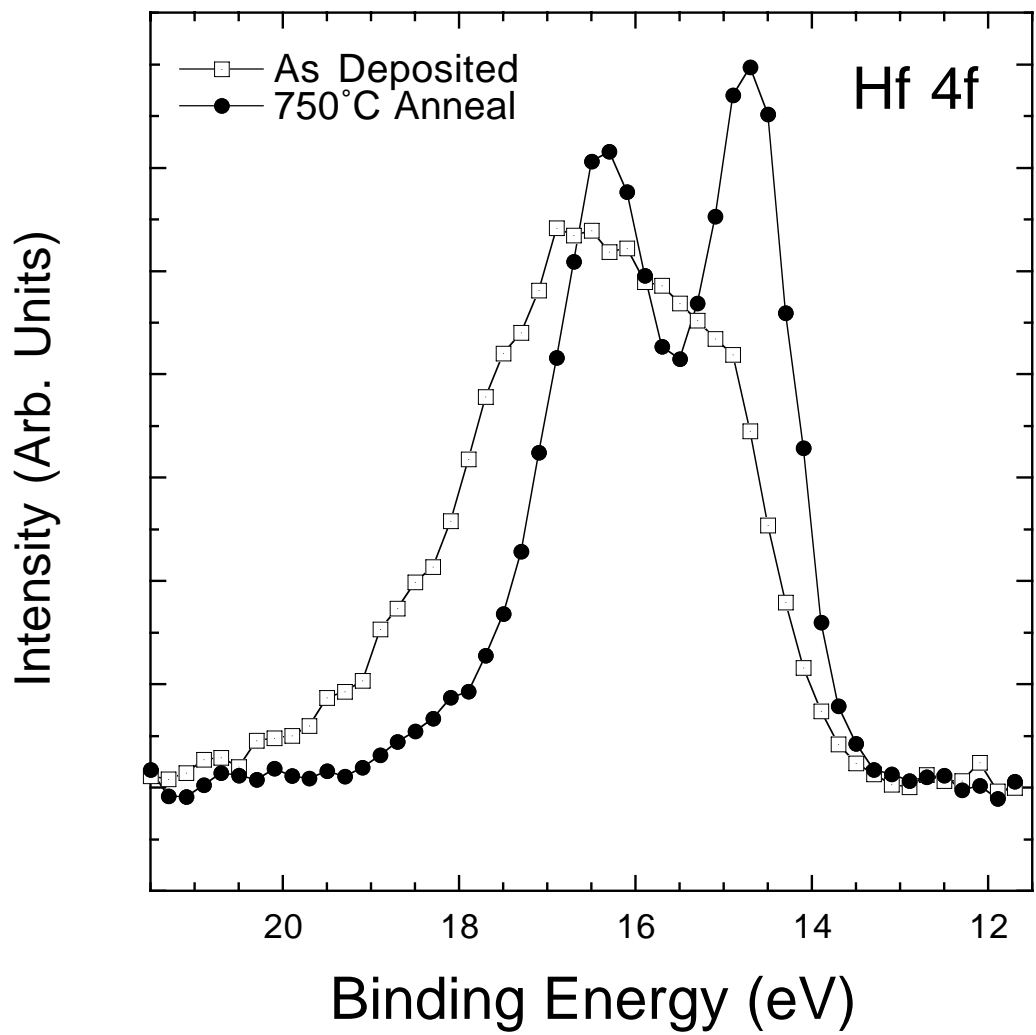

Fig. 4  Johnson-Steigelman et al.

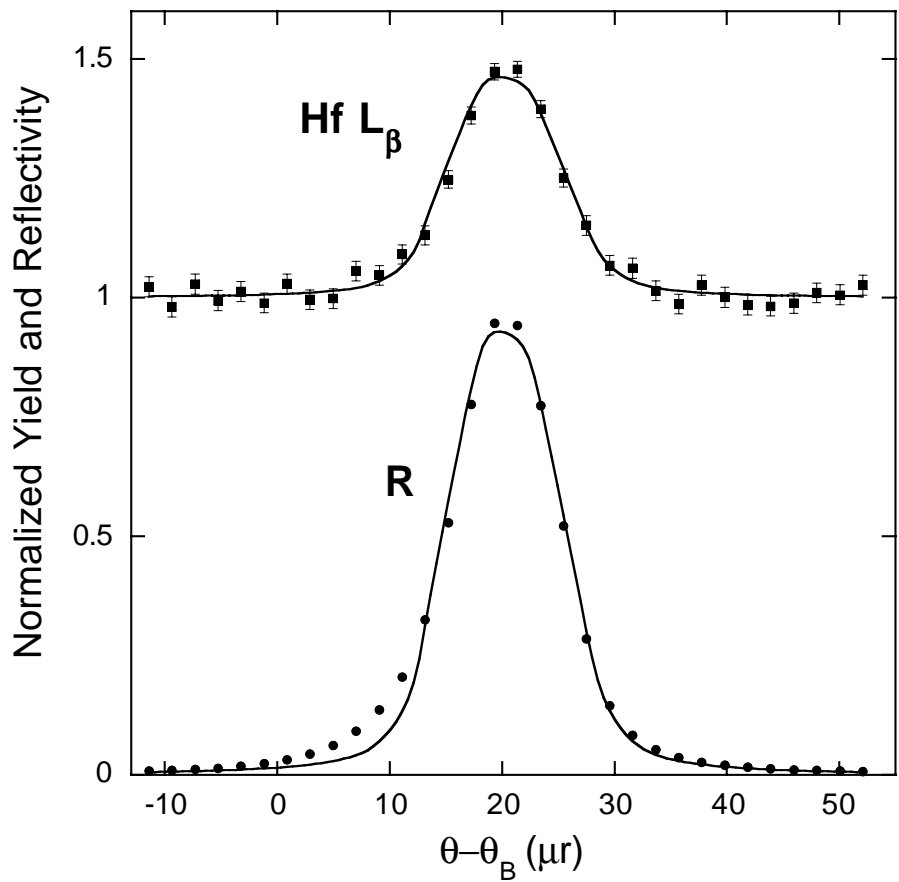

Fig. 5  Johnson-Steigelman et al.